\documentclass[conference,letterpaper]{IEEEtran}

\usepackage[utf8]{inputenc}
\usepackage[T1]{fontenc}
\usepackage{url}
\usepackage{cite}
\usepackage[cmex10]{amsmath}
\usepackage{amsfonts}
\usepackage{amssymb}
\usepackage{graphicx}
\usepackage{caption}

\newtheorem{theorem}{Theorem}
\newtheorem{lemma}{Lemma}

\DeclareMathOperator*{\argmin}{arg\,min}

\begin{document}

\title{Pipelined Gradient Coding}

\author{%
  \IEEEauthorblockN{Xian Su}
  \IEEEauthorblockA{Department of Computer Science\\
                    Florida International University\\
                    Miami, FL, USA\\
                    Email: xsu@fiu.edu}
  \and
  \IEEEauthorblockN{Jun Li}
  \IEEEauthorblockA{Queens College and The Graduate Center\\
                    City University of New York\\
                    New York, NY, USA\\
                    Email: jun.li@qc.cuny.edu}
}

\maketitle

\begin{abstract}
    In large-scale machine learning, distributed training commonly involves multiple workers evaluating the gradients of the model on different dataset partitions. A common challenge is the presence of straggling workers, which may significantly slow down training. Traditional gradient coding (GC) addresses this by duplicating dataset partitions across workers, allowing for the replacement of missing gradients from stragglers. However, GC requires workers to evaluate gradients on multiple dataset partitions in each step, potentially increasing overall training time. In this paper, we propose to pipeline GC, such that gradient evaluation is segmented across multiple steps and each worker evaluates gradients on just a single dataset partition per step. We develop the pipelined version for fractional repetition (FR) and cyclic repetition (CR), two representative dataset placement schemes in GC, and prove convergence guarantees for both. Through extensive simulations and experiments on cloud infrastructure, our schemes not only significantly reduce training time but also accelerate convergence compared to GC and other baselines.
\end{abstract}

\section{Introduction}
The training of machine-learning models is typically based on solving an optimization problem, where the model is iteratively improved by updating its parameters based on the gradients evaluated over samples from a dataset. With the rapid expansion in the size of modern models, the datasets used in the training also become larger and larger. Hence, it has become a common practice to train the model in parallel on multiple {\em workers}, where each worker evaluates the gradients over just a single dataset partition, and the gradients evaluated on different workers are aggregated on a \textit{master} to update the model~\cite{dean2012large, dekel2012optimal, zinkevich2010parallelized, Yu2017, Lee2018a, Tandon}.

\begin{figure}[t]
\centering
\includegraphics[width=\linewidth]{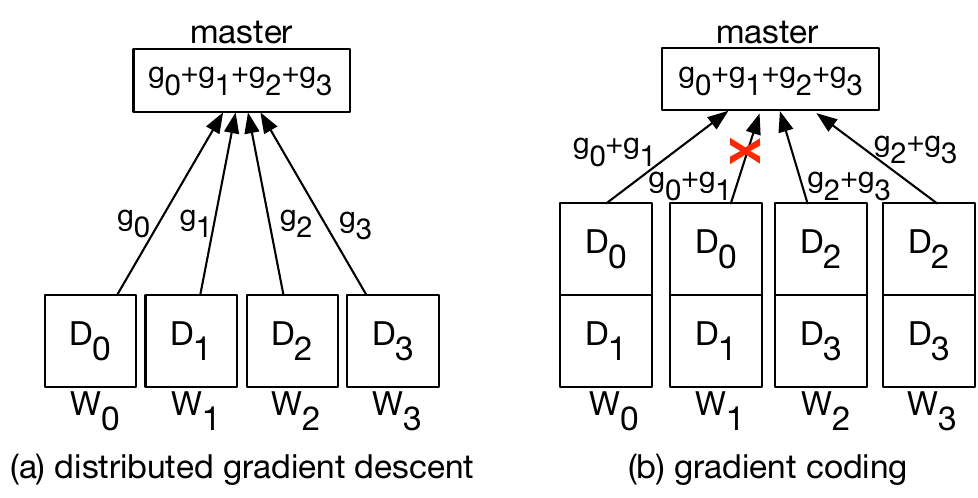}
\captionsetup{skip=2pt}
\caption{A comparison between distributed gradient descent and gradient coding.}
\label{fig:example}
\end{figure}

Fig.~\ref{fig:example}a illustrates an example of distributed gradient descent (DGD), where the dataset $D$ is partitioned into $D_0,\ldots,D_3$, and the gradients are evaluated on these four partitions as $g_0, \ldots,g_3$, respectively. The master then can aggregate them into $\sum_{i=0}^{3}g_i$ to update the model for the next step.

During the distributed training over a large number of workers, it is common or even inevitable to have some worker(s) performing worse than others, even if they have homogeneous configurations~\cite{dean2008mapreduce}. Such workers are typically known as {\em stragglers}, which can be a major bottleneck for the master to obtain the aggregated gradients and significantly delay the overall progress of the training, as each step in training requires the aggregation of gradients from {\em all} dataset partitions~\cite{dean2013tail, Tandon, li2014communication, Lee2018a, dean2008mapreduce}.

As gradients evaluated on some dataset partitions may be delayed by corresponding stragglers, we can mitigate stragglers by duplicating dataset partitions on multiple workers, such that missing gradients from one dataset partition on one straggling worker can be replaced by those evaluated on another worker.

Therefore, gradient coding (GC)~\cite{Tandon} was proposed to tolerate up to $s$ stragglers by assigning $c=s+1$ dataset partitions on each worker such that each dataset partition has $c$ replicas on different workers. As shown in Fig.~\ref{fig:example}b, each worker evaluates gradients over $c=2$ dataset partitions and uploads a linear combination of such gradients to the master. The master can then recover the aggregated gradients without coded gradients from $c-1=s$ workers.

Although GC can recover the aggregated gradients from a subset of all workers, workers are also required to evaluate gradients on multiple dataset partitions~\cite{Tandon}. It may take $c$ times more time to evaluate the gradients on all $c$ dataset partitions on each worker~\cite{ozfatura2020straggler}. If the delay incurred by the straggler is lower than the additional time incurred by the $c$ dataset partitions, the overall time needed may be even higher. In Sec.~\ref{sec:background}, we will demonstrate empirical results showing that GC can often lead to even higher completion time.

In this paper, we propose pipelined gradient coding (PGC)\footnote{This work also appears in the first author's Ph.D. dissertation~\cite{su2024thesis}.}, which mitigates stragglers without incurring the additional computational workload that GC does. As evaluating gradients on multiple dataset partitions makes GC slow, PGC breaks such computation into multiple steps in a pipelined fashion, such that each worker needs to evaluate gradients on just one of its dataset partitions in each step. Meanwhile, gradients on different dataset partitions are evaluated in sequential steps, and coded gradients can then be encoded from gradients evaluated in current and previous steps on different dataset partitions. In this way, PGC can achieve the same computational overhead as DGD while efficiently mitigating stragglers as GC.

We design PGC for two representative schemes of dataset placement in GC, {\em i.e.}, fractional repetition and cyclic repetition, and prove their convergence properties. Through extensive simulations and experiments, we demonstrate that PGC can significantly save the training time and also speed up the convergence.\footnote{This is the extended version of the paper accepted at the IEEE Information Theory Workshop (ITW) 2026.}

\section{Related Work}

Since stragglers can significantly increase tail latency in distributed training, mitigating them is critical~\cite{dean2008mapreduce}. Replication-based methods, in which the master assigns the same task to multiple workers, offer a straightforward solution~\cite{dean2008mapreduce, wang2015using, wang2019efficient, WangDa2015, ghare2004improving}, but require prohibitive extra resources to tolerate even a limited number of stragglers. Coding-based methods instead partition and encode the dataset into unique coded tasks per worker, so the master can recover the full result by decoding coded results from a subset of workers~\cite{Lee2018a, Yu2017, Yu2020a, Dutta2019, Soto2019, su2022local, ferdinand2016anytime, yang2017coded, ozfatura2021coded, mallick2020rateless}. However, as these are designed for matrix multiplication, they apply only to linear operations.

GC applies coding directly to gradients, so it extends to any model that is updated iteratively from gradients~\cite{Tandon}; the cost is added per-worker computation that can lengthen training. Beyond the original construction, a deterministic Reed-Solomon-based construction minimizes the expected computation time by optimally choosing the scheme parameters~\cite{Halbawi2018}, and Ye and Abbe proposed a communication-computation efficient GC achieving a tradeoff between computation and communication~\cite{Ye2018c}. A series of methods encode gradients into multiple vectors rather than a single one to exploit stragglers' partial results~\cite{ozfatura2019speeding, Ozfatura2018, ozfatura2021coded, Wang2021}. All of these, however, still require each worker to evaluate gradients on multiple partitions per step~\cite{ozfatura2020straggler}. Most closely related, Sequential GC~\cite{krishnan2022sequential} also exploits the temporal dimension, but through multi-round scheduling with relaxed per-task deadlines and adaptive reattempts across rounds, while each worker still evaluates $c$ partitions per task. In contrast, PGC pipelines the per-partition computation \emph{within a single} training process, so that each worker evaluates only one partition per step.

Another line of work mitigates stragglers by approximately recovering the aggregated gradients. Ignore-straggler SGD (IS-SGD) simply disregards gradients on stragglers, as the model can still converge with partially aggregated gradients~\cite{ozfatura2019distributed, Dutta2018, hanna2020adaptive, chen2016revisiting}. Approximate GC (AGC) and stochastic GC (SGC) recover more gradients than IS-SGD for the same number of stragglers and reduce per-worker overhead~\cite{Wang2019a, Charles2017a, Wang2019c, ozfatura2021coded, Bitar2019, Raviv2020}, but their code constructions sacrifice accuracy and are limited to specific combinations of parameters and straggler counts. Ignore-straggler GC (IS-GC) improves flexibility by maximizing the recovered gradients in GC for an arbitrary number of stragglers~\cite{suarbitrary}. These straggler-ignoring techniques reduce GC's training time but only partially recover the aggregated gradients, so the model converges more slowly than with fully aggregated gradients~\cite{Tandon}; moreover, all except IS-SGD still incur the additional per-worker workload. In contrast, PGC approximately recovers the fully aggregated gradients while reducing per-worker computation to the same level as DGD.

\section{Preliminary and Motivation}
\label{sec:background}

Consider a dataset $D$ with $d$ samples and $p$ features. We aim to find optimal parameters via $\beta^{*} = \argmin_{\beta\in\mathbb{R}^p} F(\beta; D) = \argmin_{\beta\in\mathbb{R}^p} \frac{1}{d}\sum_{i=0}^{d-1} f(\beta; x_i, y_i)$. In the $t$-th step, $\beta^{(t+1)}=\beta^{(t)} - \eta g^{(t)}$, where $\eta$ is the learning rate. For distributed gradient descent (DGD) with $n$ workers, $g^{(t)}=\frac{1}{n}\sum_{i=0}^{n-1} g_i^{(t)}$, where $g_i^{(t)}$ denotes the gradients on partition $D_i$ at step $t$.

\begin{figure}[t]
\centering
\includegraphics[width=\linewidth]{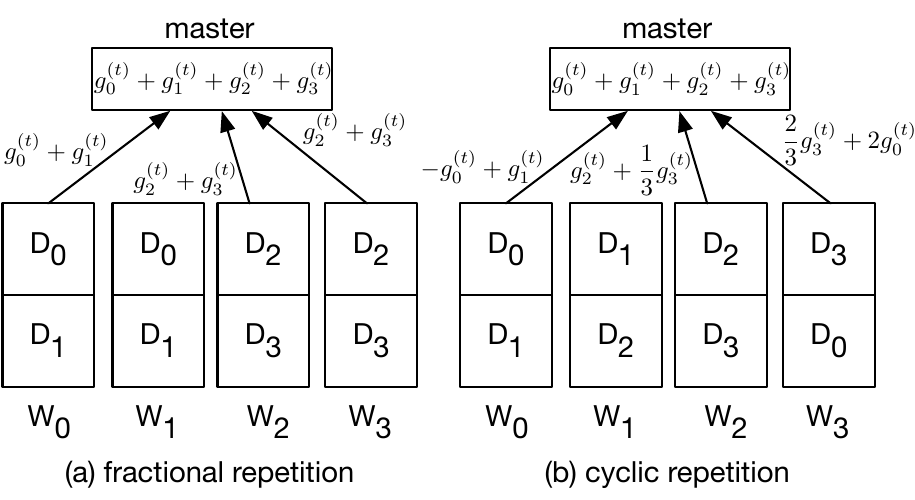}
\captionsetup{skip=2pt}
\caption{Illustrative examples of FR and CR, both with $n=4$ and $c=2$.}
\label{fig:GC}
\end{figure}

GC recovers $\sum_{i=0}^{n-1} g_i^{(t)}$ from any $n-s$ workers by assigning each worker $c=s+1$ dataset partitions. Two representative placement schemes are fractional repetition (FR) and cyclic repetition (CR), illustrated in Fig.~\ref{fig:GC}.

FR requires $c|n$, dividing $n$ workers into $\frac{n}{c}$ groups in which all workers share the same placement. Worker $W_i$ holds partitions $D_{jc},\ldots,D_{jc+(c-1)}$, where $j=\lfloor\frac{i}{c}\rfloor$; for example, with $n=4$ and $c=2$ (Fig.~\ref{fig:GC}a), $W_0,W_1$ store $D_0,D_1$ and $W_2,W_3$ store $D_2,D_3$. Each worker uploads the sum of its gradients; since workers in the same group send identical coded gradients, any one worker per group suffices, tolerating up to $s=c-1$ stragglers.

CR does not require $c|n$, placing partitions cyclically: $W_i$ holds $\{D_{j\bmod n}|j=i,\ldots,i+c-1\}$. Each worker uploads a linear combination of its $c$ partition gradients; the master applies straggler-pattern-dependent decoding coefficients to the $n-s$ received coded gradients to recover $\sum_i g_i^{(t)}$ exactly~\cite{Tandon}. This offers greater flexibility than FR at the cost of a more involved coding construction.

\begin{figure}[t]
\centering
\includegraphics[width=\linewidth]{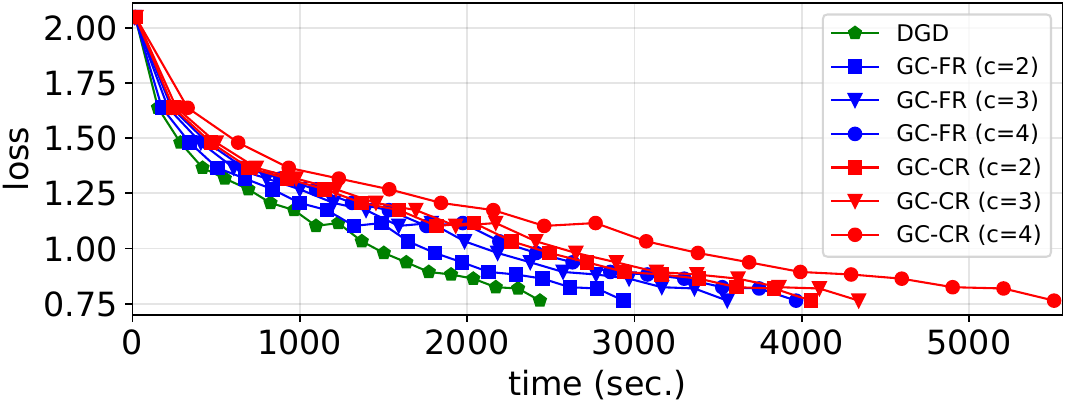}
\captionsetup{skip=2pt}
\caption{Training time per step normalized by DGD when training ResNet-18 on CIFAR-10 with $n=12$ workers on Google Cloud, showing that GC consistently requires more time than DGD due to its $c$-fold computational overhead.}
\label{fig:motivation}
\end{figure}

However, GC requires each worker to evaluate gradients on its $c = s+1$ partitions per step, multiplying per-step computation by $c$. To validate this empirically, we train a ResNet-18 model~\cite{he2016deep} on CIFAR-10~\cite{krizhevsky2009learning} for one epoch on $n=12$ Google Cloud virtual machines, comparing DGD against GC with both FR and CR. As Fig.~\ref{fig:motivation} shows, although all methods achieve the same loss per step (both recover the full gradient), GC's extra computation makes it consistently slower per step than DGD ($>100\%$); GC-FR is faster than GC-CR owing to its lower coding complexity, but both lag behind DGD, and the gap widens as $s$ grows. This exposes a paradox: unless stragglers are sufficiently frequent and slow, a mechanism meant to accelerate training by mitigating stragglers actually prolongs it, leaving total training time no better, or even worse, than DGD.

\begin{figure}[t]
\centering
\includegraphics[width=.615\linewidth]{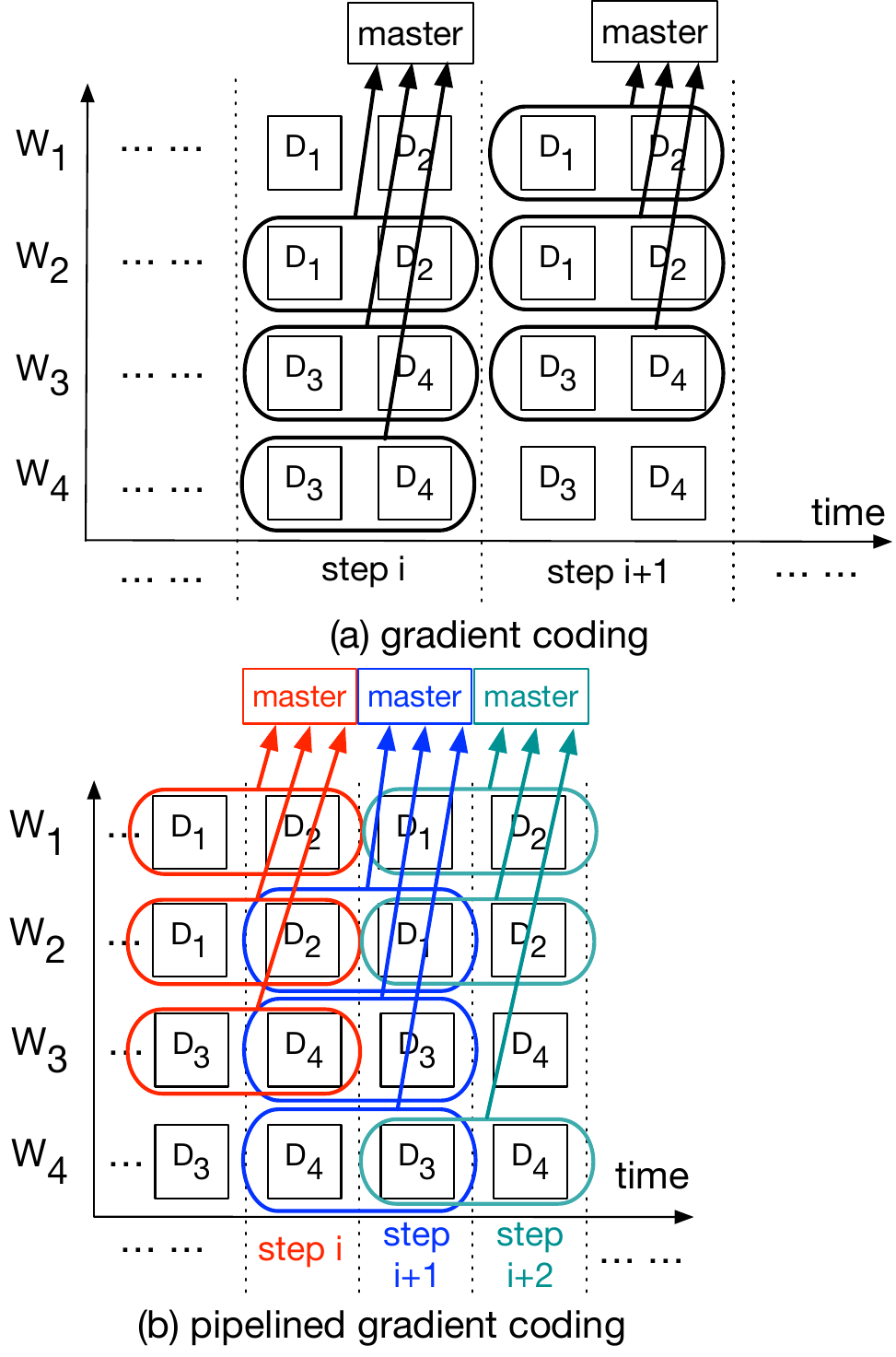}
\captionsetup{skip=2pt}
\caption{A comparison between gradient coding and pipelined gradient coding.}
\label{fig:intuition}
\end{figure}

Fig.~\ref{fig:intuition} illustrates how PGC addresses this. Rather than evaluating $c$ partitions in one step, each worker rotates through its assigned partitions one per step, retaining the most recent gradient from each. At each step, it assembles a coded gradient by combining the fresh gradient with $c-1$ retained stale values, replicating the GC coded gradient structure; the master then collects responses from the $n-s$ fastest workers and decodes as in GC. Each worker still uploads exactly one coded gradient per step, so PGC's per-step communication is identical to GC and DGD and incurs no additional communication overhead, with only the local computation pipelined. The cost is that $c-1$ component gradients are stale, introducing a bounded approximation error that diminishes with the learning rate. Sections~\ref{sec:FR} and~\ref{sec:CR} detail the pipelined encoding and decoding of PGC for FR and CR, respectively, and analyze their convergence, establishing that PGC converges geometrically under standard assumptions, matching GC's convergence structure while reducing per-step computation to that of DGD.

\section{Pipelined Gradient Coding with \\ Fractional Repetition}
\label{sec:FR}

In this section, we present the design of PGC for FR. The design for CR is presented in Section~\ref{sec:CR}.

\subsection{Coding Scheme}

With FR, the $n$ dataset partitions are distributed so that worker $W_i$ holds the $c$ partitions $\{D_{jc},\ldots,D_{jc+(c-1)}\}$, where $j=\lfloor\frac{i}{c}\rfloor$. In GC, at each step every worker receives the current model parameters from the master, evaluates the gradients on \emph{all} $c$ of its partitions, and uploads their sum as the coded gradient $\sum_{l=0}^{c-1} g^{(t)}_{jc+l}$. As long as at least one worker per group is non-straggling, the master can aggregate the full gradient and update the model, tolerating up to $s=c-1$ stragglers, at the cost of evaluating $c$ partitions per worker at every step.

PGC-FR reduces this to one partition per step. Each worker still receives the model parameters at every step, but evaluates the gradients on only one of its $c$ partitions, rotating through them cyclically: at step $t$, worker $W_i$ evaluates $D_{jc+((t-1)\bmod c)}$. Its other $c-1$ partitions are therefore not evaluated at step $t$; to still form a coded gradient, PGC replaces the missing gradients by the most recent {\em stale} gradients evaluated within the previous $c-1$ steps, assuming they do not differ much from those that would be evaluated now. The resulting coded gradient $\hat{g}_i^{(t)} = \sum_{l=0}^{c-1} g^{(t-l)}_{jc+((t-1-l)\bmod c)}$ coincides with the GC coded gradient $\sum_{l=0}^{c-1} g^{(t)}_{jc+l}$ when all evaluations are current, and approximates it when the staleness is small.

Once the master has received coded gradients from at least $n-s$ workers, it recovers the (approximate) aggregated gradient exactly as in GC by ignoring the staleness: it selects one coded gradient per group and sums them across groups. Since there are not enough evaluated partitions before step $c$, we add an initial step $t=0$ at which each worker evaluates all $c$ of its partitions on the initial parameters to warm up the pipeline; because partitions are placed exactly as in GC, GC itself can be applied at this step if stragglers must be tolerated.

\subsection{Convergence Analysis}
\label{sec:FR_convergence}

We now analyze the convergence of PGC-FR. Let $s(t,i)\in[t-c+1,t]$ denote the step at which dataset partition $D_i$'s gradient contributing to the update at step $t$ was evaluated. The update rule is $\beta^{(t+1)}=\beta^{(t)} - \eta \hat{g}^{(t)}$, where $\hat{g}^{(t)}=\frac{1}{n}\sum_{i=0}^{n-1} g_i^{(s(t,i))}$. We establish convergence through two lemmas and Theorem~\ref{theorem:FR}.

\begin{lemma}
\label{lemma:FR1}
Assume $\mathbb{E}[g^{(t)}_i] = \nabla F(\beta^{(t)})$ for all $i$. Then:
\begin{align*}
&\mathbb{E}[||\nabla F(\beta^{(t)}) - \hat{g}^{(t)}||^2_2] \\
&\quad = \mathbb{E}[||\nabla F(\beta^{(t)}) - \nabla F(\beta^{(s(t,i))})||^2_2] \\
&\qquad - \mathbb{E}[||\nabla F(\beta^{(s(t,i))})||^2_2] + \mathbb{E}[||\hat{g}^{(t)}||^2_2].
\end{align*}
\end{lemma}

Lemma~\ref{lemma:FR1} decomposes the gradient estimation error into a staleness term (controlled by $\gamma$) and a second-moment term (controlled by $\xi$). It measures the difference between the true gradient $\nabla F(\beta^{(t)})$ and the coded estimate $\hat{g}^{(t)}$.

\begin{lemma}
\label{lemma:FR2}
Assume $\mathbb{E}[||g_i^{(s(t,i))}-\nabla F(\beta^{(s(t,i))})||^2_2] \leq \xi\mathbb{E}[||\nabla F(\beta^{(s(t,i))})||^2_2]$ for some $\xi\in[0,1)$. Then:
\[
\mathbb{E}\!\left[\left|\left|\textstyle\sum_{i=0}^{n-1}g_i^{(s(t,i))}\right|\right|^2_2\right] \leq (n + \xi)\textstyle\sum_{i=0}^{n-1}\mathbb{E}[||\nabla F(\beta^{(s(t,i))})||^2_2].
\]
\end{lemma}

Lemma~\ref{lemma:FR2} bounds the squared norm of the aggregated gradient by $(n+\xi)$ times the sum of individual squared norms; the factor $n$ arises because cross terms between distinct unbiased gradient evaluations vanish, and $\xi$ captures the stochastic gradient noise.

\begin{theorem}
\label{theorem:FR}
Assume that $F(\beta)$ is $\lambda$-strongly convex and $\nabla F(\beta)$ is $L$-Lipschitz continuous. Assume the gradient variance is bounded: $\mathbb{E}[||g_i^{(s(t,i))}-\nabla F(\beta^{(s(t,i))})||^2_2] \leq \xi\mathbb{E}[||\nabla F(\beta^{(s(t,i))})||^2_2]$ for some $\xi\in[0,1)$. Assume staleness is bounded: $\mathbb{E}[||\nabla F(\beta^{(t)}) - \nabla F(\beta^{(s(t,i))})||^2_2] \leq \gamma \mathbb{E}[||\nabla F(\beta^{(t)})||^2_2]$ for some $\gamma\in[0,1)$. Then with learning rate $\eta \leq \frac{1}{nL(n+\xi)}$:
\[
\mathbb{E}[F(\beta^{(t)})] - F(\beta^*) \leq \left(1-\frac{\lambda(1-\gamma)}{nL(n+\xi)}\right)^t\!\!(F(\beta^{(0)}) - F(\beta^*)).
\]
\end{theorem}

Theorem~\ref{theorem:FR} guarantees geometric convergence under standard assumptions. The staleness is inherently bounded by the pipeline depth ($s(t,i)\in[t-c+1,t]$), so $\gamma$ is governed by $c$ and the learning rate $\eta$, vanishing as $\eta\to0$. The complete proofs of Lemmas~\ref{lemma:FR1} and~\ref{lemma:FR2} and Theorem~\ref{theorem:FR} are given in Appendices~\ref{app:FR1}, \ref{app:FR2}, and~\ref{app:FR}, respectively.

\section{Pipelined Gradient Coding with \\ Cyclic Repetition}
\label{sec:CR}

\subsection{Motivation}

\begin{figure}[t]
\centering
\includegraphics[width=\linewidth]{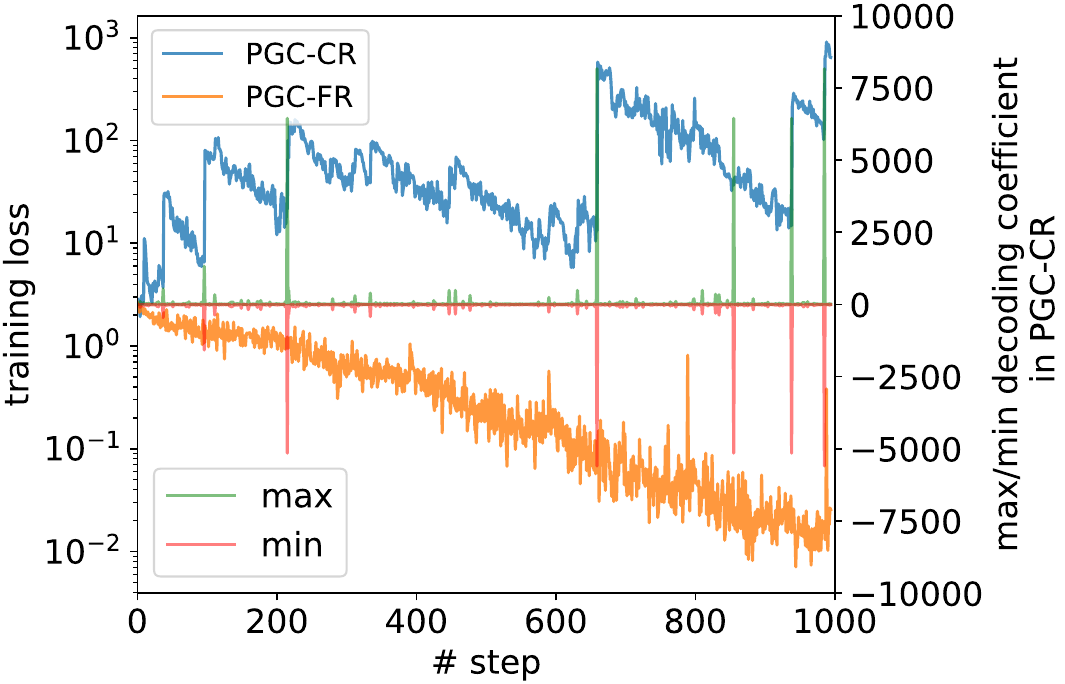}
\captionsetup{skip=2pt}
\caption{Training with PGC-FR and PGC-CR for $n=12$, $c=6$. The right axis shows the maximum and minimum decoding coefficients $\alpha_{m,k}$ of the na\"{i}ve PGC-CR design.}
\label{fig:CR2}
\end{figure}

Similar to PGC-FR, our initial attempt to build PGC-CR was to directly apply the pipelining principle to CR using the coding scheme from GC. This attempt, however, failed. As shown in Fig.~\ref{fig:CR2}, this na\"{i}ve design does not converge, while PGC-FR under the same training configuration converges stably.

The failure stems from how GC-CR decodes. In GC-CR, each worker $W_j$ uploads a {\em weighted} sum of its $c$ partition gradients, $\hat{g}_j^{(t)} = \sum_{l=0}^{c-1} c_{j,l}\, g^{(t)}_{(j+l)\bmod n}$, with fixed encoding coefficients $c_{j,l}$ (for GC-FR, all $c_{j,l}=1$). To recover $\sum_i g_i^{(t)}$ from the $n-s$ non-straggling workers $J^{(t)}=\{j_0,\ldots,j_{n-s-1}\}$, the master solves a linear system for straggler-pattern-dependent decode weights $d_m(J^{(t)})$ and computes
\begin{align*}
\sum_{m=0}^{n-s-1} d_m(J^{(t)})\,\hat{g}_{j_m}^{(t)} &= \sum_{m=0}^{n-s-1} d_m(J^{(t)})\sum_{l=0}^{c-1} c_{j_m,l}\, g^{(t)}_{(j_m+l)\bmod n} \\
&= \sum_{k=0}^{n-1} g_k^{(t)}.
\end{align*}
The effective weight applied to partition $k$'s gradient contributed by worker $m$ is thus the \emph{decoding coefficient} $\alpha_{m,k} = d_m(J^{(t)})\, c_{j_m,l}$ with $(j_m+l)\bmod n = k$. Exact recovery requires $\sum_m \alpha_{m,k}=1$ for each $k$, but the individual $\alpha_{m,k}$ are unconstrained in magnitude: for poorly conditioned straggler patterns they reach the thousands, up to $\pm 7500$ for $n=12$, $c=6$. In GC-FR, by contrast, each partition is contributed by exactly one worker with coefficient $1$.

\begin{figure}[t]
\centering
\includegraphics[width=.72\linewidth]{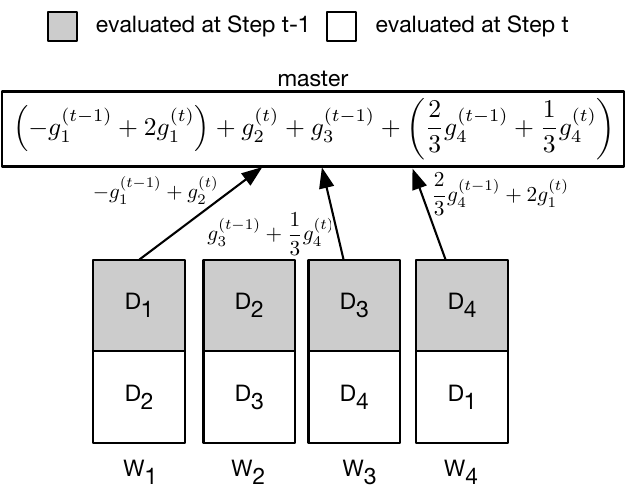}
\captionsetup{skip=2pt}
\caption{The na\"{i}ve PGC-CR construction for $n=4$, $c=2$.}
\label{fig:CR}
\end{figure}

Fig.~\ref{fig:CR} illustrates the na\"{i}ve PGC-CR construction for $n=4$, $c=2$: at step $t$ each worker evaluates its second partition and reuses the first, evaluated at step $t-1$. Decoding recovers, for instance, partition $D_1$'s gradient as $-g_1^{(t-1)}+2g_1^{(t)}$, {\em i.e.}, with decoding coefficients $\{-1,2\}$ that sum to $1$. When both copies are current this yields $-g_1^{(t)}+2g_1^{(t)}=g_1^{(t)}$ exactly; with staleness, however, the stale copy $g_1^{(t-1)}$ is weighted by $-1$, so its staleness error is magnified rather than averaged out. Even in this small example a decoding coefficient already reaches magnitude $2$, and for larger $n$ and $c$ it can grow without bound.

With PGC's stale gradients, each contribution to partition $k$ carries a staleness error $\delta_{m,k}$. In PGC-FR this error enters with coefficient $1$ and stays small; in na\"{i}ve PGC-CR each $\delta_{m,k}$ is scaled by $|\alpha_{m,k}|$, and since contributions come from different prior steps the errors do not cancel despite $\sum_m\alpha_{m,k}=1$. The net error is amplified by up to $7500\times$. Fig.~\ref{fig:CR2} plots the maximum and minimum decoding coefficients $\alpha_{m,k}$ at each step alongside the training loss: the na\"{i}ve PGC-CR loss spikes precisely when these decoding coefficients spike.

\subsection{Coding Scheme}

We therefore design PGC-CR to avoid large decoding coefficients. Rather than requiring exact recovery of the full gradient sum, we relax to a {\em weighted} sum whose expected coefficient for each partition equals $1$, {\em i.e.}, $\mathbb{E}\!\left[\frac{1}{\mu}\left|N_k^{(t)}\right|\right]=1$, where $\mu=\frac{(n-s)c}{n}$ and $|N_k^{(t)}|$ is the total number of times a gradient on partition $D_k$ appears across the $n-s$ received coded gradients at step $t$. This target is achievable: since $\sum_{k=0}^{n-1} |N_k^{(t)}| = (n-s)c$ and each worker is equally likely to straggle, $\mathbb{E}[|N_k^{(t)}|]=\frac{(n-s)c}{n}=\mu$, so $\mathbb{E}[\frac{1}{\mu}|N_k^{(t)}|]=1$.

With $D_{(j+l)\bmod n}$ on $W_j$, $l=0,\ldots,c-1$, worker $W_j$ evaluates $D_{(j+(t-1)\bmod c)\bmod n}$ at step $t$ and forms the coded gradient $\hat{g}_j^{(t)} = \sum_{l=0}^{c-1} g^{(t-l)}_{(j+(t-1-l)\bmod c)\bmod n}$ from the most recent $c$ evaluations, all with coefficient $1$. The master collects coded gradients from the $n-s$ fastest workers and computes $\hat{g}^{(t)} = \frac{1}{n\mu}\sum_{m=0}^{n-s-1}\hat{g}_{j_m}^{(t)}$, where the additional factor $\frac{1}{n}$ rescales the recovered weighted sum to match DGD's averaged gradient.

Let $s(t,i,j)\in[t-c+1,t]$ denote the step at which the copy of partition $D_i$'s gradient held by its $j$-th holder ($0\leq j\leq c-1$) and used at step $t$ was evaluated; $s(t,i,j)<t$ when that copy is stale. We now show that this estimator is unbiased, over the straggler pattern, for the averaged stale-gradient sum $\bar{g}^{(t)}=\frac{1}{nc}\sum_{i=0}^{n-1}\sum_{j=0}^{c-1} g^{(s(t,i,j))}_i$. Since each of the $n$ partitions is placed on exactly $c$ workers, summing the coded gradients of \emph{all} $n$ workers reproduces each partition's gradient once per holder,
\begin{equation}
\textstyle\sum_{j=0}^{n-1}\hat{g}_j^{(t)}=\sum_{i=0}^{n-1}\sum_{j=0}^{c-1} g^{(s(t,i,j))}_i,
\label{eq:cr-fullsum}
\end{equation}
which is independent of the straggler pattern, as each worker still evaluates and stores its own gradients regardless of whether it straggles. As each worker is equally likely to straggle, it is among the received $n-s$ with probability $\frac{n-s}{n}$, so $\mathbb{E}[\sum_{m=0}^{n-s-1}\hat{g}_{j_m}^{(t)}]=\frac{n-s}{n}\sum_{j=0}^{n-1}\hat{g}_j^{(t)}$. Combining with~\eqref{eq:cr-fullsum} and using $\frac{1}{n\mu}=\frac{1}{(n-s)c}$,
\begin{equation}
\mathbb{E}[\hat{g}^{(t)}]=\frac{1}{n\mu}\cdot\frac{n-s}{n}\sum_{i,j} g^{(s(t,i,j))}_i=\frac{1}{nc}\sum_{i,j} g^{(s(t,i,j))}_i=\bar{g}^{(t)},
\label{eq:cr-unbiased}
\end{equation}
consistent with the estimator analyzed in Theorem~\ref{theorem:CR}. This design also ensures $0 \leq \frac{1}{\mu}|N_k^{(t)}| < 2$. The lower bound is immediate. For the upper bound, $|N_k^{(t)}|\leq\min\{c,\,n-s\}$, since each worker holds at most $c$ partitions and only $n-s$ coded gradients are received. As $c=s+1$: when $s\leq\frac{n-1}{2}$, $c\leq n-s$ and $\frac{1}{\mu}|N_k^{(t)}|\leq\frac{c}{\mu}=\frac{n}{n-s}<2$; when $s>\frac{n-1}{2}$, $\frac{1}{\mu}|N_k^{(t)}|\leq\frac{n-s}{\mu}=\frac{n}{s+1}<2$. Hence no partition contributes more than twice its intended weight in any step, so the amplification of stale-gradient errors stays bounded, in stark contrast to the up-to-$7500\times$ amplification of the na\"{i}ve design.

\subsection{Convergence Analysis}
\label{sec:CR_convergence}

We now analyze the convergence of PGC-CR. Recall from~\eqref{eq:cr-unbiased} that, over the straggler pattern, $\hat{g}^{(t)}$ is unbiased for $\bar{g}^{(t)} = \frac{1}{nc}\sum_{i=0}^{n-1}\sum_{j=0}^{c-1}g^{(s(t,i,j))}_i$. Let $\bar{\nabla}^{(t)} = \frac{1}{nc}\sum_{i=0}^{n-1}\sum_{j=0}^{c-1}\nabla F(\beta^{(s(t,i,j))})$ be the corresponding averaged full gradient. The proof of Theorem~\ref{theorem:CR} rests on two lemmas.

\begin{lemma}
\label{lemma:CR1}
Assume $\mathbb{E}[g^{(t)}_i] = \nabla F(\beta^{(t)})$ for all $i$. Then:
\begin{align*}
\mathbb{E}[||\nabla F(\beta^{(t)}) - \hat{g}^{(t)}||^2_2] &= \mathbb{E}[||\nabla F(\beta^{(t)}) - \bar{\nabla}^{(t)}||^2_2] \\
&\quad - \mathbb{E}[||\bar{\nabla}^{(t)}||^2_2] + \mathbb{E}[||\bar{g}^{(t)}||^2_2].
\end{align*}
\end{lemma}

Lemma~\ref{lemma:CR1} decomposes the gradient estimation error into a staleness bias term ($\mathbb{E}[||\nabla F(\beta^{(t)}) - \bar{\nabla}^{(t)}||^2_2]$, controlled by $\gamma$) and a stochastic variance term ($\mathbb{E}[||\bar{g}^{(t)}||^2_2]$, controlled by $\xi$), with a negative cross term that tightens the bound. The proof adds and subtracts $\bar{\nabla}^{(t)}$ and uses unbiasedness to cancel cross terms.

\begin{lemma}
\label{lemma:CR2}
Assume $\mathbb{E}[||g^{(s(t,i,j))}_i - \nabla F(\beta^{(s(t,i,j))})||^2_2] \leq \xi\mathbb{E}[||\nabla F(\beta^{(s(t,i,j))})||^2_2]$ for some $\xi\in[0,1)$. Then:
\begin{align*}
&\mathbb{E}\!\left[\!\left|\!\left|\textstyle\sum_{i=0}^{n-1}\sum_{j=0}^{c-1}g^{(s(t,i,j))}_i\right|\!\right|^2_2\right] \\
&\quad \leq (nc+\xi)\textstyle\sum_{i=0}^{n-1}\sum_{j=0}^{c-1}\mathbb{E}\!\left[\!\left|\!\left|\nabla F(\beta^{(s(t,i,j))})\right|\!\right|^2_2\right].
\end{align*}
\end{lemma}

Lemma~\ref{lemma:CR2} bounds the squared norm of the total gradient sum by $(nc+\xi)$ times the sum of squared individual norms. The factor $nc$ arises because, for any two distinct evaluations $(i_1,j_1)\neq(i_2,j_2)$, the cross term $\mathbb{E}[(g^{(s)}_{i_1}-\nabla F(\beta^{(s)}))^{\!\top}\!(g^{(s)}_{i_2}-\nabla F(\beta^{(s)}))]$ vanishes: each stochastic gradient is an unbiased, conditionally independent estimate given its evaluation point, so the expectation factorizes to zero. Only the $nc$ diagonal terms survive, and the variance bound $\xi$ controls each residual. The complete derivation is given in Appendix~\ref{app:CR2}.

\begin{theorem}
\label{theorem:CR}
Assume that $F(\beta)$ is $\lambda$-strongly convex and $\nabla F(\beta)$ is $L$-Lipschitz continuous. Assume that for some $\xi\in[0,1)$, $\mathbb{E}[||g^{(s(t,i,j))}_{i}-\nabla F(\beta^{(s(t,i,j))})||^2_2] \leq \xi \mathbb{E}[||\nabla F(\beta^{(s(t,i,j))})||^2_2]$. Define $\bar{\nabla}^{(t)} = \frac{1}{nc}\sum_{i=0}^{n-1}\sum_{j=0}^{c-1}\nabla F(\beta^{(s(t,i,j))})$. Assume that for some $\gamma\in[0,1)$:
$\mathbb{E}\!\left[\left|\left|\nabla F(\beta^{(t)})-\bar{\nabla}^{(t)}\right|\right|^2_2\right] \leq \gamma \mathbb{E}[||\nabla F(\beta^{(t)})||^2_2]$.
Then with learning rate $\eta \leq \frac{1}{L(nc+\xi)}$:
\[
\mathbb{E}[F(\beta^{(t)})] - F(\beta^*) \leq \left(1-\frac{\lambda(1-\gamma)}{L(nc+\xi)}\right)^t\!\!(F(\beta^{(0)}) - F(\beta^*)).
\]
\end{theorem}

Theorem~\ref{theorem:CR} guarantees geometric convergence for PGC-CR. The complete proofs of Lemmas~\ref{lemma:CR1} and~\ref{lemma:CR2} and Theorem~\ref{theorem:CR} are given in Appendices~\ref{app:CR1}, \ref{app:CR2}, and~\ref{app:CR}, respectively.

Unlike PGC-FR, where each partition's gradient comes from one worker, PGC-CR averages gradients from different workers with varying staleness; this mixing reduces the adverse effect of staleness and speeds up convergence in practice.

\section{Evaluation}
\label{sec:evaluation}

\subsection{Implementation}
\label{subsec:implementation}
We implement PGC using \textsf{Ray}~\cite{ray} for distributed computing and \textsf{PyTorch} for training. Each worker evaluates gradients on one dataset partition per step and keeps the gradients evaluated in the previous $c-1$ steps; it then encodes the latest gradient together with these stale gradients into a coded gradient and sends it to the master. The master collects coded gradients from the $n-s$ fastest workers using \textsf{Ray.wait()}, treats the remaining $s$ workers as stragglers, decodes using the respective FR or CR scheme, updates the model, and broadcasts the new parameters for the next step. The optimizer is \textsf{torch.optim.SGD} so that gradients can be extracted and updated directly; DGD is the special case with $c=1$ and $s=0$.

We compare PGC against DGD, GC (with FR and CR), IS-SGD, and IS-GC~\cite{suarbitrary}. In IS-SGD, each worker evaluates one dataset partition and uploads its gradient, and the master aggregates the $n-s$ fastest (equivalent to DGD when $s=0$). With GC, each worker evaluates its $c$ partitions per step and encodes them before uploading. IS-GC combines GC with straggler ignoring so that it needs a smaller $c$ to tolerate the same number of stragglers. To ensure a fair comparison, we control all random seeds so that gradients evaluated on the same dataset partition are identical across schemes and all models start from the same initial parameters.

\subsection{Simulation}

We first evaluate through simulations on a local high-performance computing (HPC) cluster with $n=12$ workers, training ResNet-18 on ImageNet~\cite{chrabaszcz2017downsampled}. To simulate stragglers, we add independent random delays on each worker following exponential distributions with means of $1$ and $2$ seconds, respectively. Results are averaged over $5$ runs; training runs for $25$ epochs.

\begin{figure}[t]
\centering
\includegraphics[width=\linewidth]{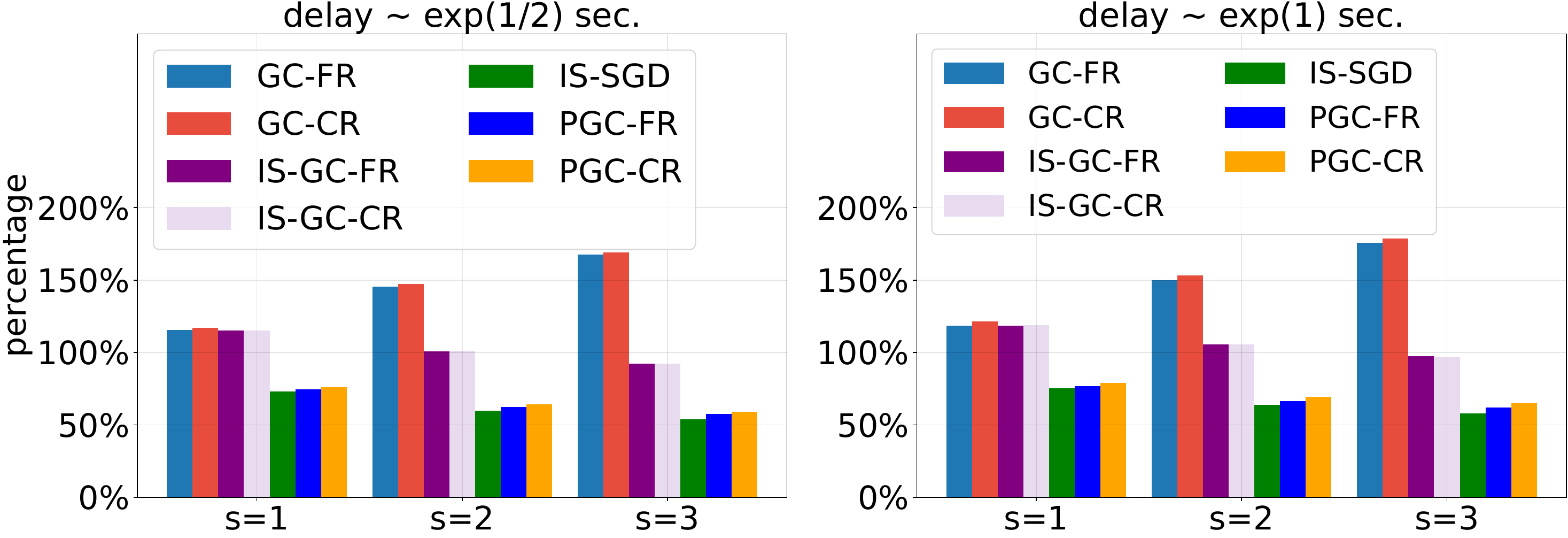}
\captionsetup{skip=2pt}
\caption{Time per step when training ResNet-18 on ImageNet, normalized by DGD ($100\%$). PGC achieves stable near-DGD per-step time regardless of $s$.}
\label{fig:simulationBar}
\end{figure}

\textbf{Time per step.}
We vary $s$ from $1$ to $3$, so GC and PGC use $c=s+1\in\{2,3,4\}$, while DGD and IS-SGD use $c=1$ and IS-GC fixes $c=2$ (its lowest possible value). Unlike DGD, which cannot tolerate stragglers and is therefore delayed by the slowest worker, every other scheme completes a step once the master receives (coded) gradients from the $n-s$ fastest workers. Fig.~\ref{fig:simulationBar} normalizes per-step time against DGD ($100\%$). GC (FR and CR) exceeds $100\%$ because each worker computes on $c=s+1$ partitions; a larger $c$ adds more computational overhead, which eventually outweighs the time saved from stragglers. IS-GC lowers per-step time at $s=2,3$ thanks to its smaller $c$. In contrast, PGC stays stable and near-DGD regardless of $s$, since each worker always evaluates exactly one partition per step while still tolerating up to $s$ stragglers.

\begin{figure}[t]
\centering
\includegraphics[width=.70\linewidth]{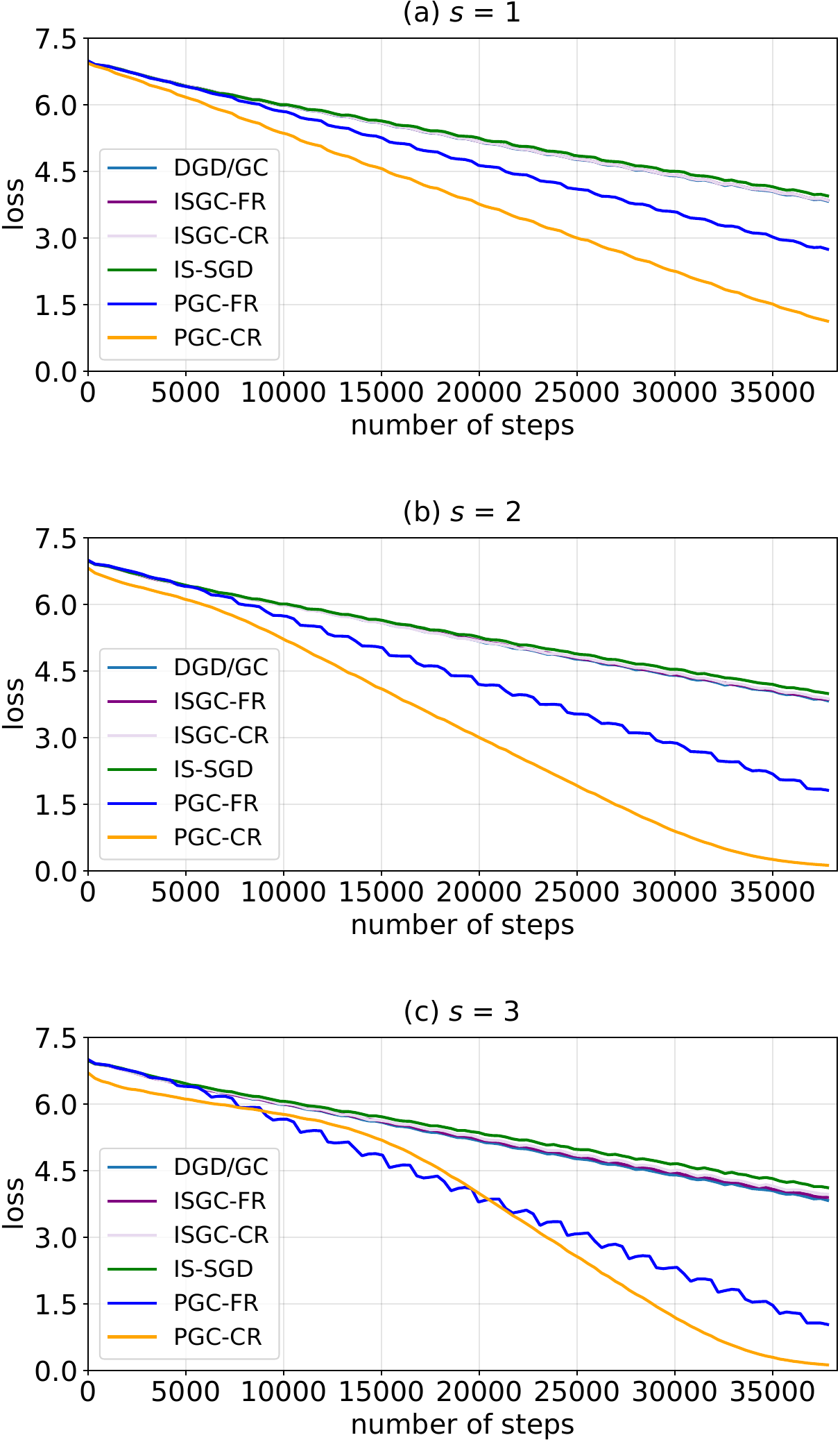}
\captionsetup{skip=2pt}
\caption{Training loss over steps for ResNet-18 on ImageNet with $s=1,2,3$. GC/DGD curves coincide; PGC converges faster with the advantage growing for larger $s$.}
\label{fig:simulationCurve}
\end{figure}

\textbf{Convergence.}
Fig.~\ref{fig:simulationCurve} shows training loss over $25$ epochs ($38{,}150$ steps). GC coincides with DGD since both recover the same aggregated gradient, and IS-SGD, IS-GC, DGD, and GC all converge at very similar rates: IS-SGD discards straggler gradients, reducing the effective batch size, while IS-GC avoids this bias but its $c$-fold per-step computation offsets the straggler-tolerance benefit. PGC converges noticeably faster, and the advantage grows with $s$. This shows that stale gradients approximate current-step gradients well; in fact, mild staleness can even accelerate convergence, consistent with observations in asynchronous learning~\cite{Dutta2018}. It is somewhat surprising that PGC-CR consistently outperforms PGC-FR for every $s$. We attribute this to CR's ``mixed'' staleness: in PGC-FR each partition's gradient comes from a single worker and thus carries a single staleness, whereas PGC-CR averages coded gradients across workers, so gradients on the same partition arrive with different staleness. Since staler gradients tend to deviate more from the correct descent direction, mixing them dilutes this adverse effect (further quantified in Sec.~\ref{sec:cloud}).

\subsection{Experiments in the Cloud}
\label{sec:cloud}

We train ResNet-18~\cite{he2016deep} on CIFAR-10~\cite{krizhevsky2009learning} on a Google Cloud cluster of $n=12$ workers (\textsf{n2-highmem-2}) and one master (\textsf{e2-highmem-2}), with $s\in\{1,2,3\}$. No artificial delays are added; natural stragglers arise from the cloud environment. Training runs until loss reaches $0.5$; each scheme is run $10$ times. Batch size is $256$ and learning rate is $0.01$.

\begin{figure}[t]
\centering
\includegraphics[width=.7\linewidth]{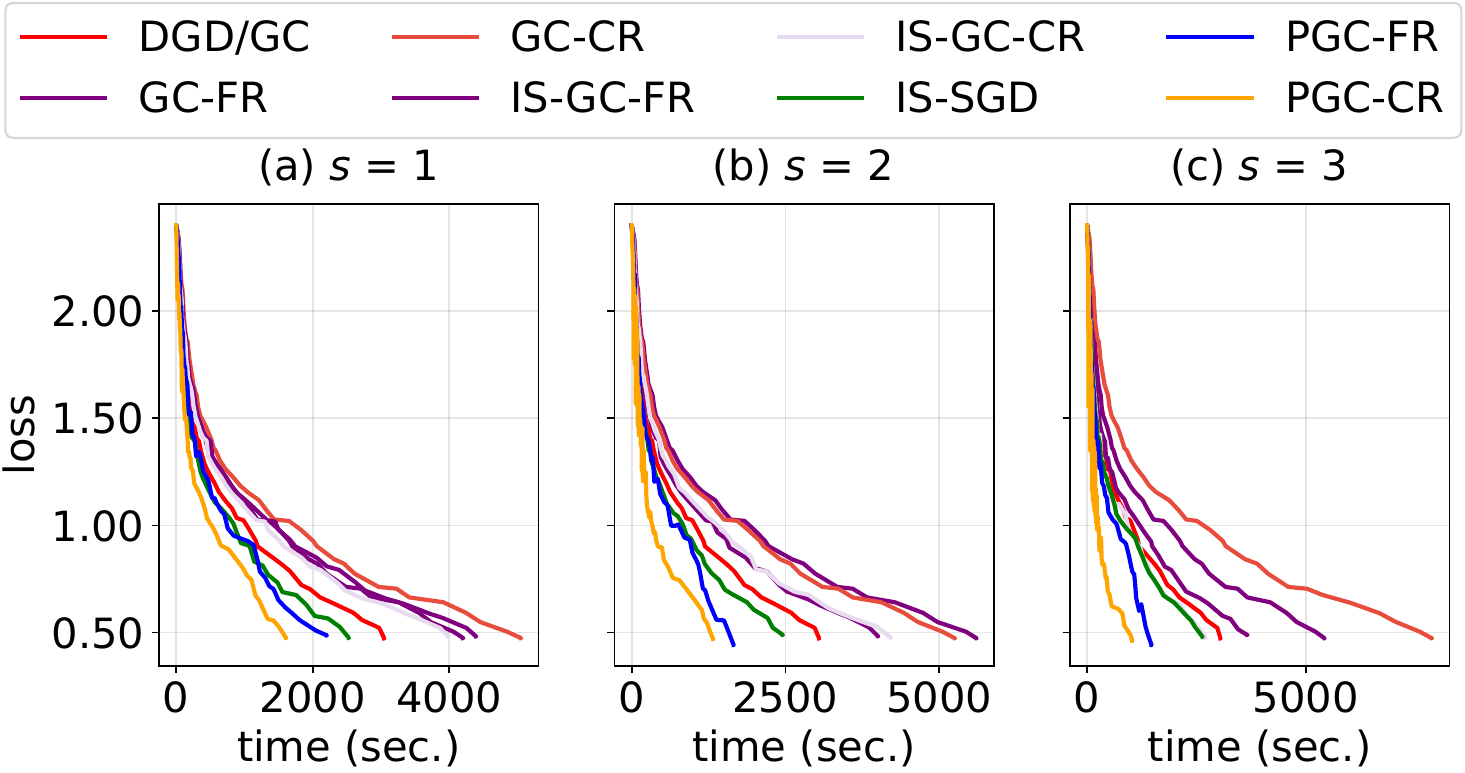}
\captionsetup{skip=2pt}
\caption{Loss reduction over time when training ResNet-18 on CIFAR-10 using DGD, GC, IS-GC, IS-SGD, and PGC for $s=1,2,3$.}
\label{fig:mainCurve}
\end{figure}

\begin{figure}[t]
\centering
\includegraphics[width=\linewidth]{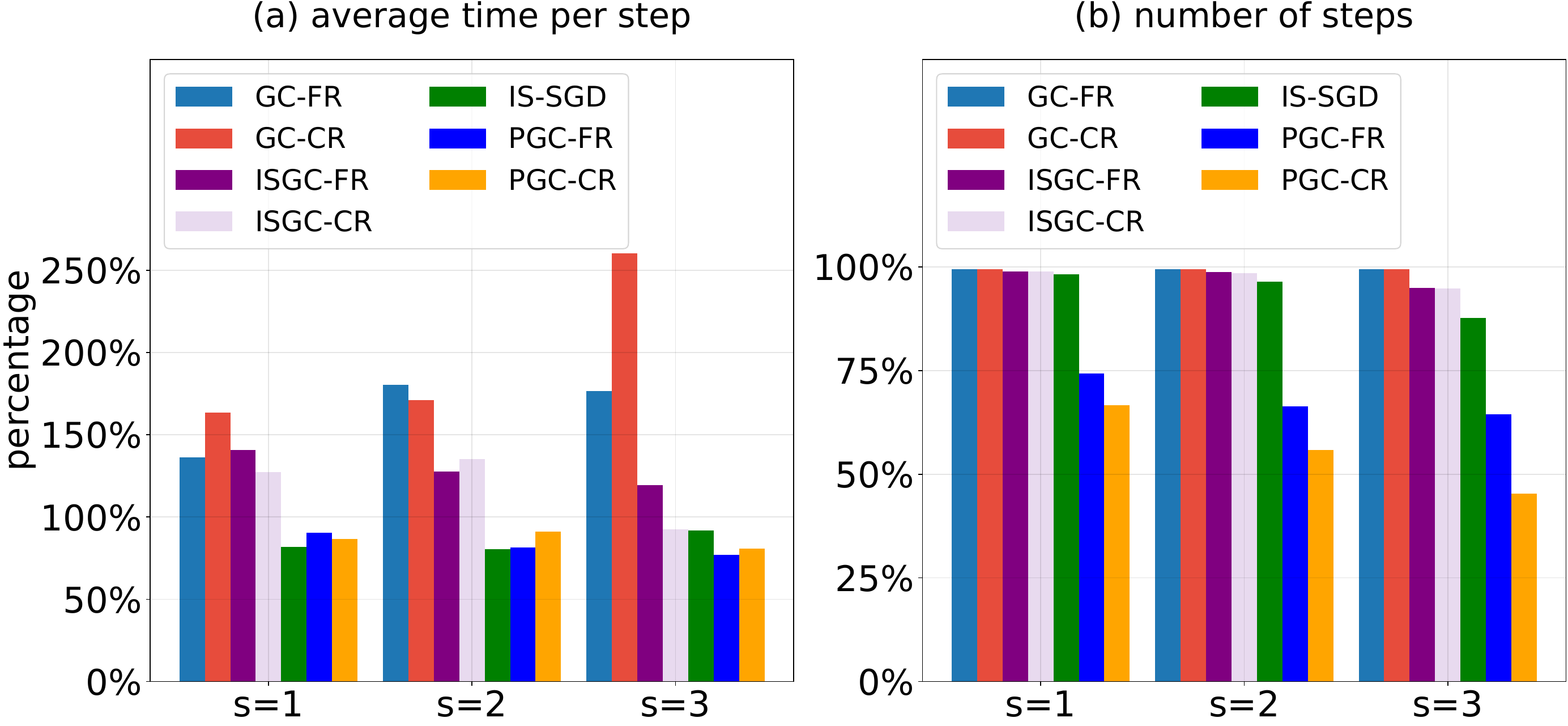}
\captionsetup{skip=2pt}
\caption{Time consumption training ResNet-18 on CIFAR-10, normalized by DGD ($100\%$).}
\label{fig:mainBar}
\end{figure}

\textbf{Training time.}
Fig.~\ref{fig:mainCurve} shows how the loss decreases over training time. As in the simulations, GC (FR and CR) takes the longest for each $s$ due to its higher computational overhead. IS-GC beats DGD only at $s=3$, where it tolerates more stragglers at a relatively low $c$; IS-SGD is slightly better than DGD at $s=1$ and saves more as $s$ grows. Regardless of $s$, PGC reaches the loss threshold in the least time, with PGC-CR faster than PGC-FR in all cases, owing to its better convergence rate, as the simulations suggest. To explain this, we analyze the average time per step and the number of steps separately below.

\textbf{Time per step.}
Fig.~\ref{fig:mainBar}a compares average time per step, normalized to DGD ($100\%$). GC consumes more time than DGD due to its computational overhead, while IS-GC costs less at higher $s$. Since PGC evaluates only one partition per step, it achieves almost the same time per step as IS-SGD.

\textbf{Convergence.}
Fig.~\ref{fig:mainBar}b shows steps to reach the loss threshold. We still use DGD as the baseline ($100\%$). GC always matches DGD as it recovers the same aggregated gradients. PGC achieves significantly fewer steps, up to $54.85\%$ fewer than DGD.
For IS-GC, we fix $c=2$ for all $s$, the minimum for straggler ignoring, so its per-step computation is twice DGD's; despite this, IS-GC beats DGD only at $s=3$, since the tolerated-straggler fraction must exceed the extra-computation fraction to yield a gain. IS-SGD instead discards straggler gradients, avoiding the overhead but degrading convergence through a downward bias on the effective batch size. PGC uses one partition per step (like IS-SGD) while recovering coded gradients that approximate the full aggregated gradient (like GC), attaining the best of both.

As in simulations, PGC-CR converges faster than PGC-FR, mirroring our analysis: the contraction factor $\left(1-\frac{\lambda(1-\gamma)}{L(nc+\xi)}\right)$ in Theorem~\ref{theorem:CR} is smaller than $\left(1-\frac{\lambda(1-\gamma)}{nL(n+\xi)}\right)$ in Theorem~\ref{theorem:FR} whenever $c \leq n$. Intuitively, CR's ``mixed'' staleness, {\em i.e.}, gradients on the same partition coming from different workers with different delays, averages out the adverse effect of stale gradients, akin to variance reduction in asynchronous optimization~\cite{Dutta2018}. Combining better per-step time and faster convergence, PGC substantially reduces total training time.

\section{Conclusion}

We proposed pipelined gradient coding (PGC), which overcomes the $c$-fold computation overhead of gradient coding by pipelining: each worker evaluates only one partition per step and reuses stale gradients to form coded gradients. We proved convergence for both FR and CR variants and showed that PGC matches DGD's efficiency while outperforming prior schemes in training time.

\bibliographystyle{IEEEtran}
\bibliography{main}

\appendices

\section{Proof of Lemma~\ref{lemma:FR1}}
\label{app:FR1}

Completing the square in $\mathbb{E}[||\nabla F(\beta^{(t)}) - \hat{g}^{(t)}||^2_2]$ by adding and subtracting $\nabla F(\beta^{(s(t,i))})$,
\begin{align*}
\mathbb{E}[||\nabla F(\beta^{(t)}) - \hat{g}^{(t)}||^2_2] &= \mathbb{E}[||\nabla F(\beta^{(t)}) - \nabla F(\beta^{(s(t,i))})||^2_2] \\
&\quad + \mathbb{E}[||\hat{g}^{(t)} - \nabla F(\beta^{(s(t,i))})||^2_2] \\
&\quad - 2\mathbb{E}[(\nabla F(\beta^{(t)}) - \nabla F(\beta^{(s(t,i))}))^{\!\top}\\
&\qquad\quad (\hat{g}^{(t)} - \nabla F(\beta^{(s(t,i))}))].
\end{align*}
Because $\mathbb{E}[g_i^{(s(t,i))}]=\nabla F(\beta^{(s(t,i))})$, we have $\mathbb{E}[\hat{g}^{(t)} - \nabla F(\beta^{(s(t,i))})]=0$, so the last term vanishes; moreover $\mathbb{E}[||\hat{g}^{(t)} - \nabla F(\beta^{(s(t,i))})||^2_2] = \mathbb{E}[||\hat{g}^{(t)}||^2_2] - \mathbb{E}[||\nabla F(\beta^{(s(t,i))})||^2_2]$. Substituting yields the claim. \hfill$\blacksquare$

\section{Proof of Lemma~\ref{lemma:FR2}}
\label{app:FR2}

Expanding $\mathbb{E}[||\sum_{i}(g^{(s(t,i))}_{i} - \nabla F(\beta^{(s(t,i))}))||^2_2]$, every cross term $\mathbb{E}[(g^{(s(t,i_1))}_{i_1} - \nabla F(\beta^{(s(t,i_1))}))^{\!\top}(g^{(s(t,i_2))}_{i_2} - \nabla F(\beta^{(s(t,i_2))}))]$ with $i_1\neq i_2$ vanishes, since the two factors are independent and each has zero mean. Hence
$\mathbb{E}[||\sum_{i}(g^{(s(t,i))}_{i} - \nabla F(\beta^{(s(t,i))}))||^2_2] = \sum_{i}\mathbb{E}[||g^{(s(t,i))}_{i} - \nabla F(\beta^{(s(t,i))})||^2_2] \leq \xi\sum_{i}\mathbb{E}[||\nabla F(\beta^{(s(t,i))})||^2_2]$.
On the other hand, using unbiasedness the same quantity equals $\mathbb{E}[||\sum_{i} g^{(s(t,i))}_{i}||^2_2] - \mathbb{E}[||\sum_{i}\nabla F(\beta^{(s(t,i))})||^2_2]$. Combining the two and bounding $\mathbb{E}[||\sum_{i}\nabla F(\beta^{(s(t,i))})||^2_2]\leq n\sum_{i}\mathbb{E}[||\nabla F(\beta^{(s(t,i))})||^2_2]$ gives the claim. \hfill$\blacksquare$

\section{Proof of Theorem~\ref{theorem:FR}}
\label{app:FR}

Recall that for PGC-FR the update is $\beta^{(t+1)}=\beta^{(t)}-\eta\hat{g}^{(t)}$ with $\hat{g}^{(t)}=\frac{1}{n}\sum_{i=0}^{n-1} g^{(s(t,i))}_{i}$. By $L$-Lipschitz continuity of $\nabla F$,
$F(\beta^{(t+1)}) \leq F(\beta^{(t)}) + \nabla F(\beta^{(t)})^{\!\top}(\beta^{(t+1)} -\beta^{(t)}) + \frac{L}{2}||\beta^{(t+1)} - \beta^{(t)}||^2_2$.
Substituting $\beta^{(t+1)} = \beta^{(t)} -\eta\hat{g}^{(t)}$, applying the polarization identity, and taking expectations,
\begin{align*}
\mathbb{E}[F(\beta^{(t+1)})] &\leq \mathbb{E}[F(\beta^{(t)})] - \tfrac{\eta}{2}\mathbb{E}[||\nabla F(\beta^{(t)})||^2_2] \\
&\quad - \tfrac{\eta}{2n}\textstyle\sum_{i}\mathbb{E}[||g^{(s(t,i))}_{i}||^2_2] \\
&\quad + \tfrac{\eta}{2n}\textstyle\sum_{i}\mathbb{E}[||g^{(s(t,i))}_{i} - \nabla F(\beta^{(t)})||^2_2] \\
&\quad + \tfrac{L\eta^2}{2}\mathbb{E}[||\hat{g}^{(t)}||^2_2].
\end{align*}
Applying Lemma~\ref{lemma:FR1}, the staleness bound $\mathbb{E}[||\nabla F(\beta^{(t)}) - \nabla F(\beta^{(s(t,i))})||^2_2]\leq\gamma\mathbb{E}[||\nabla F(\beta^{(t)})||^2_2]$, and Lemma~\ref{lemma:FR2},
\begin{align*}
\mathbb{E}[F(\beta^{(t+1)})] &\leq \mathbb{E}[F(\beta^{(t)})] - \tfrac{\eta(1-\gamma)}{2}\mathbb{E}[||\nabla F(\beta^{(t)})||^2_2] \\
&\quad - \tfrac{\eta}{2n}\!\left(1 - \tfrac{L\eta(n+\xi)}{n}\right)\!\textstyle\sum_{i}\mathbb{E}[||\nabla F(\beta^{(s(t,i))})||^2_2].
\end{align*}
Since $\eta \leq \frac{1}{nL(n+\xi)}$, the last term is non-positive, so
$\mathbb{E}[F(\beta^{(t+1)})] \leq \mathbb{E}[F(\beta^{(t)})] - \frac{1-\gamma}{2nL(n+\xi)}\mathbb{E}[||\nabla F(\beta^{(t)})||^2_2]$.
By $\lambda$-strong convexity, $||\nabla F(\beta)||^2_2 \geq 2\lambda(F(\beta)-F(\beta^*))$, hence
$\mathbb{E}[F(\beta^{(t+1)})] - F(\beta^*)\leq (1-\frac{\lambda(1-\gamma)}{nL(n+\xi)})(\mathbb{E}[F(\beta^{(t)})]-F(\beta^{*}))$.
Applying this contraction recursively from step $0$,
\[
\mathbb{E}[F(\beta^{(t)})] - F(\beta^*)\leq \left(1-\tfrac{\lambda(1-\gamma)}{nL(n+\xi)}\right)^t\!(F(\beta^{(0)})-F(\beta^{*})),
\]
which is the stated bound. \hfill$\blacksquare$

\section{Proof of Lemma~\ref{lemma:CR1}}
\label{app:CR1}

Recall from~\eqref{eq:cr-unbiased} in Sec.~\ref{sec:CR} that, in expectation over the straggler pattern, $\mathbb{E}[\hat{g}^{(t)}]=\bar{g}^{(t)}=\frac{1}{nc}\sum_{i=0}^{n-1}\sum_{j=0}^{c-1}g^{(s(t,i,j))}_{i}$.

Following the reasoning of Appendix~\ref{app:FR1}, expand $\mathbb{E}[||\nabla F(\beta^{(t)}) - \hat{g}^{(t)}||^2_2]$ by adding and subtracting $\bar{\nabla}^{(t)}=\frac{1}{nc}\sum_{i,j}\nabla F(\beta^{(s(t,i,j))})$. Since $\mathbb{E}[\hat{g}^{(t)}]=\bar{g}^{(t)}$ has mean $\bar{\nabla}^{(t)}$ under the unbiasedness assumption, the cross term vanishes, giving
\begin{align*}
\mathbb{E}[||\nabla F(\beta^{(t)}) - \hat{g}^{(t)}||^2_2] &= \mathbb{E}[||\nabla F(\beta^{(t)}) - \bar{\nabla}^{(t)}||^2_2] \\
&\quad + \mathbb{E}[||\bar{\nabla}^{(t)} - \hat{g}^{(t)}||^2_2].
\end{align*}
Expanding the second term and using $\mathbb{E}[\hat{g}^{(t)}]=\bar{g}^{(t)}$,
$\mathbb{E}[||\bar{\nabla}^{(t)} - \hat{g}^{(t)}||^2_2] = \mathbb{E}[||\bar{g}^{(t)}||^2_2] - \mathbb{E}[||\bar{\nabla}^{(t)}||^2_2]$.
Substituting completes the proof:
\begin{align*}
\mathbb{E}[||\nabla F(\beta^{(t)}) - \hat{g}^{(t)}||^2_2] &= \mathbb{E}[||\nabla F(\beta^{(t)}) - \bar{\nabla}^{(t)}||^2_2] \\
&\quad - \mathbb{E}[||\bar{\nabla}^{(t)}||^2_2] + \mathbb{E}[||\bar{g}^{(t)}||^2_2]. \tag*{$\blacksquare$}
\end{align*}

\section{Proof of Lemma~\ref{lemma:CR2}}
\label{app:CR2}

Expand $\mathbb{E}[||\sum_{i=0}^{n-1}\sum^{c-1}_{j=0}(g^{(s(t,i,j))}_{i} - \nabla F(\beta^{(s(t,i,j))}))||^2_2]$. Each of the cross terms
$\mathbb{E}[(g^{(s(t,i_1,j_1))}_{i_1} - \nabla F(\beta^{(s(t,i_1,j_1))}))^{\!\top}(g^{(s(t,i_2,j_2))}_{i_2} - \nabla F(\beta^{(s(t,i_2,j_2))}))]$
with $(i_1,j_1)\neq(i_2,j_2)$ vanishes: each factor has zero conditional mean by unbiasedness, and distinct evaluations are independent, so the expectation factorizes to $0$. Therefore
\begin{align*}
&\mathbb{E}\!\left[\left|\left|\textstyle\sum_{i,j}(g^{(s(t,i,j))}_{i} - \nabla F(\beta^{(s(t,i,j))}))\right|\right|^2_2\right] \\
&\quad = \textstyle\sum_{i,j}\mathbb{E}[||g^{(s(t,i,j))}_{i} - \nabla F(\beta^{(s(t,i,j))})||^2_2].
\end{align*}
Using unbiasedness, the same quantity equals $\mathbb{E}[||\sum_{i,j}g^{(s(t,i,j))}_{i}||^2_2] - \mathbb{E}[||\sum_{i,j}\nabla F(\beta^{(s(t,i,j))})||^2_2]$. Combining the two expressions, applying the variance bound $\xi$, and bounding $\mathbb{E}[||\sum_{i,j}\nabla F(\beta^{(s(t,i,j))})||^2_2] \leq nc\sum_{i,j}\mathbb{E}[||\nabla F(\beta^{(s(t,i,j))})||^2_2]$ yields
\begin{equation*}
\mathbb{E}\!\left[\left|\left|\textstyle\sum_{i,j}g^{(s(t,i,j))}_{i}\right|\right|^2_2\right] \leq (nc+\xi)\textstyle\sum_{i,j}\mathbb{E}[||\nabla F(\beta^{(s(t,i,j))})||^2_2]. \tag*{$\blacksquare$}
\end{equation*}

\section{Proof of Theorem~\ref{theorem:CR}}
\label{app:CR}

By $L$-Lipschitz continuity, as in Appendix~\ref{app:FR},
$F(\beta^{(t+1)}) \leq F(\beta^{(t)}) + \nabla F(\beta^{(t)})^{\!\top}(\beta^{(t+1)} -\beta^{(t)}) + \frac{L}{2}||\beta^{(t+1)} - \beta^{(t)}||^2_2$.
Substituting $\beta^{(t+1)} = \beta^{(t)} -\eta\hat{g}^{(t)}$, applying the polarization identity, and taking expectations,
\begin{align*}
\mathbb{E}[F(\beta^{(t+1)})] &\leq \mathbb{E}[F(\beta^{(t)})] - \tfrac{\eta}{2}\mathbb{E}[||\nabla F(\beta^{(t)})||^2_2] \\
&\quad - \tfrac{\eta}{2}\mathbb{E}[||\hat{g}^{(t)}||^2_2] + \tfrac{\eta}{2}\mathbb{E}[||\hat{g}^{(t)} - \nabla F(\beta^{(t)})||^2_2] \\
&\quad + \tfrac{L\eta^2}{2}\mathbb{E}[||\hat{g}^{(t)}||^2_2].
\end{align*}
Applying Lemma~\ref{lemma:CR1} and the staleness bound $\mathbb{E}[||\nabla F(\beta^{(t)})-\bar{\nabla}^{(t)}||^2_2]\leq\gamma\mathbb{E}[||\nabla F(\beta^{(t)})||^2_2]$,
\begin{align*}
\mathbb{E}[F(\beta^{(t+1)})] &\leq \mathbb{E}[F(\beta^{(t)})] - \tfrac{\eta(1-\gamma)}{2}\mathbb{E}[||\nabla F(\beta^{(t)})||^2_2] \\
&\quad - \tfrac{\eta}{2}\mathbb{E}[||\bar{\nabla}^{(t)}||^2_2] + \tfrac{L\eta^2}{2}\mathbb{E}[||\hat{g}^{(t)}||^2_2].
\end{align*}
Since $\mathbb{E}[||\hat{g}^{(t)}||^2_2] = \frac{1}{n^2c^2}\mathbb{E}[||\sum_{i,j}g^{(s(t,i,j))}_{i}||^2_2]$, Lemma~\ref{lemma:CR2} gives
$\mathbb{E}[||\hat{g}^{(t)}||^2_2] \leq \frac{nc+\xi}{n^2c^2}\sum_{i,j}\mathbb{E}[||\nabla F(\beta^{(s(t,i,j))})||^2_2]$. Hence
\begin{align*}
\mathbb{E}[F(\beta^{(t+1)})] &\leq \mathbb{E}[F(\beta^{(t)})] - \tfrac{\eta(1-\gamma)}{2}\mathbb{E}[||\nabla F(\beta^{(t)})||^2_2] \\
&\quad - \tfrac{\eta}{2n^2c^2}\!\left(1 - L\eta(nc+\xi)\right)\\
&\qquad\quad\cdot\textstyle\sum_{i,j}\mathbb{E}[||\nabla F(\beta^{(s(t,i,j))})||^2_2],
\end{align*}
where we used $\mathbb{E}[||\bar{\nabla}^{(t)}||^2_2]\leq\frac{1}{nc}\sum_{i,j}\mathbb{E}[||\nabla F(\beta^{(s(t,i,j))})||^2_2]$. With $\eta \leq \frac{1}{L(nc+\xi)}$ the last term is non-positive, so
$\mathbb{E}[F(\beta^{(t+1)})] \leq \mathbb{E}[F(\beta^{(t)})] - \frac{1-\gamma}{2L(nc+\xi)}\mathbb{E}[||\nabla F(\beta^{(t)})||^2_2]$.
Applying $\lambda$-strong convexity and unrolling the recursion yields
$\mathbb{E}[F(\beta^{(t)})] - F(\beta^*) \leq (1-\frac{\lambda(1-\gamma)}{L(nc+\xi)})^t(F(\beta^{(0)}) - F(\beta^*))$. \hfill$\blacksquare$

\end{document}